\documentclass[11pt]{JHEP3}
\usepackage{amsmath,amssymb}
\newcommand{\ba}{\begin{alignat}{3}}

\title{
The Conformal Transformation in General Single Field Inflation 
with Non-Minimal Coupling}

\author{
Takahiro Kubota, 
Nobuhiko Misumi, 
Wade Naylor and 
Naoya Okuda 
\\
Department of Physics, Graduate School of Science, Osaka University, 
Toyonaka, Osaka 560-0043, Japan 
\\
E-mail:
\email{kubota@cep.osaka-u.ac.jp},
\email{misumi@het.phys.sci.osaka-u.ac.jp},
\email{naylor@phys.sci.osaka-u.ac.jp},
\email{okuda@het.phys.sci.osaka-u.ac.jp},
}

\abstract
{The method of a conformal transformation is applied to a   
general class of single field inflation models
with non-minimal coupling to gravity 
and non-standard kinetic terms, 
in order to reduce the cosmological perturbative calculation 
to the conventional minimal coupling case to all orders 
in perturbation theory. Our analysis is made  simple by
 the fact that all perturbation variables in the comoving gauge 
 are conformally invariant to all orders. 
The structure of the vacuum, on which 
cosmological correlation functions are evaluated, is also 
discussed. We  show how quantization in the Jordan frame for non-minimally 
coupled inflation models can be equivalently 
implemented in the Einstein frame. It is thereafter argued that the 
 general $N$-point cosmological correlation functions  
(of the curvature perturbation) are independent of the conformal frame. }

\begin{document}
\maketitle

\section{Introduction}\label{Intro}

It has been by now well-established that inflation can naturally 
solve cosmological problems such as 
the flatness, homogeneity and monopole  problems, thereby avoiding 
 a fine-tuning for the initial condition of the history of our Universe. 
Inflation is also known to predict almost scale invariant 
Gaussian density perturbations. 
At the next stage of observation,  the detection of non-Gaussianity 
\cite{komatsu1}-\cite{komatsu3} 
is one of the central concerns, since a detailed study of the non-Gaussianity 
and the shape of the bispectrum \cite{babich}\cite{baumann} 
in particular could constrain various inflationary models. 
It has been argued that in slow roll models of a single scalar inflaton 
the non-Gaussianity is negligibly small \cite{maldacena}\cite{acquavia}.
In the last several years there has been a lot of work exploring 
inflation models in which large non-Gaussianity could be produced. 
A partial list of such models includes DBI type inflation, k-inflation, 
curvaton scenarios, ghost inflation and so on.

In the present paper we will consider a general 
 class of inflationary models 
in which a single inflaton field couples non-minimally to background 
geometry. In such a broad class of inflationary models, we can 
explore the classification of the shape of the bispectrum. 
It has been known rather well that  the method of conformal 
transformation makes it possible to put the calculation of cosmological 
correlations into the minimal coupling case, and we will actually make 
 full use of the method throughout this work. 
What we would like to add to this common wisdom 
is that the reduction to the minimal coupling case is in fact possible 
 to all orders in perturbation theory. This paves the way for 
 avoiding technical difficulties 
associated with the non-minimal coupling to a considerable extent 
and for alleviating the need for separate studies.

The method of conformal transformation has been applied before 
in cosmology to discuss a higher derivative gravity theory \cite{MFB}.
It has been proved  in 
\cite{makino}  that the comoving curvature perturbation 
on slices orthogonal to comoving worldlines 
in the Jordan and Einstein frames coincides at the first order. 
It has also been shown in \cite{sugifuta} that the curvature perturbation 
on uniform-density hypersurfaces  agrees in these two frames up to 
 second order. 
A couple of years ago Chiba and Yamaguchi \cite{chiba}
showed a very remarkable fact 
that all perturbation variables in the comoving gauge 
are conformally invariant to all orders of perturbation. 
More recently Gong et al. \cite{gong} 
have given another proof of the conformal invariance and have discussed 
its implications in the $\delta N$ formalism.
Motivated by the observation of Chiba and Yamaguchi and that of 
Gong et al. we now  go
one step further to show the usefulness of the conformal 
invariance of the perturbation and to present a coherent method 
to perform  practical calculations of cosmological correlations 
in a general conformal frame.

The usefulness of the conformal transformation  has been 
demonstrated in the context of cosmological perturbations 
 by many people 
\cite{sbb} - \cite{ds}. While we share basic ideas with these 
previous authors, we would like to emphasize that our argument can be 
extended rather easily to all orders in the cosmological perturbation 
in contrast with previous works. An all 
order argument is made possible by the fact that, due to our 
choice of gauge,  the fluctuation part is  separated from 
 the conformal transformation factor.

In connection  with our work, let us recall that Seery and Lidsey 
\cite{seerylidsey} and 
Chen et al. \cite{chen1} investigated the shape of non-Gaussianity 
in a general Lagrangian of  single field inflation models minimally 
coupled to gravity. In the case of arbitrary sound speed the shape 
is classified in terms of five parameters.
Their work has been recently extended to the non-minimal 
coupling case by Qiu and Yang \cite{Qiu:2010dk}. 
The computation of the cosmological correlation becomes necessarily difficult 
in the Jordan frame. This is particularly so 
when one has to solve equations of motion for  the mode functions 
of the  curvature perturbation. Our technique 
shows that the difficulties in the non-minimal case 
are only ostensible and that 
all of the calculations in \cite{Qiu:2010dk} can be reproduced 
by a conformal transformation from those in \cite{seerylidsey} - 
 \cite{chen1} .

 The present paper is organized as follows. After presenting the  
  general single field inflation model in Sec. \ref{sec:general}, 
  we apply the conformal 
 transformation and discuss  connections between  various quantities in 
 the Jordan and Einstein frames in Sec. \ref{sec:conformal}. 
In Sec. \ref{sec: adm} we express the quadratic and cubic effective actions 
 with respect to the  curvature perturbation in the  Einstein  frame, 
and then we  transform the action back  into the Jordan  frame in Sec. 
\ref{sec:jordan}. The connection between the calculations in 
\cite{seerylidsey} -  \cite{chen1}   
and those in  \cite{Qiu:2010dk} is clarified there. 
In Sec. \ref{sec:frame}  the connection between 
the mode-functions in Jordan and Einstein frames is discussed. 
It is argued that the vacuum, on which correlation functions are computed, is identical in the two frames.
The fact that $N$-point cosmological correlation functions agree in Jordan and 
Einstein frames 
is proved there.  Our work is summarized in Sec. \ref{sec:summary}. 

\section{General single field inflation}
\label{sec:general}

In the following sections we discuss the general models that include 
a scalar field coupled non-minimally to gravity and go on to
perform a conformal transformation to put the equations
into a form which is the same as in the minimal coupling case. 
First we consider the following generalized Einstein-scalar 
gravitational system
\begin{eqnarray}
S=\frac{1}{2}\int dt\:d^{3}\vec{x} \sqrt{-g}\Big [ \frac{1}{8\pi G}
f(\varphi )R +2P(\varphi , X) 
\Big ], 
\label{action}
\end{eqnarray}
where we have introduced the notation
\begin{eqnarray}
X=-\frac{1}{2}g^{\mu \nu}\partial _{\mu}\varphi \partial _{\nu}\varphi ~. 
\end{eqnarray}
Hereafter we omit the gravitational constant $G$ by setting $8\pi G=1$.

The functional  $f(\varphi)$ of a scalar field $\varphi$ 
coupled to the scalar curvature $R$ can be  arbitrary.  
The minimal coupling case corresponds to choosing $f(\varphi)=1$; 
however attention  has been paid in literature  \cite{sugifuta}, 
\cite{Hosotani}- \cite{yokoyama} to non-minimal cases such as 
$
f(\varphi)=1-\xi\,\varphi^{2}~, 
$
where $\xi $ is a parameter. 
We wish, however,  to proceed most generally without specifying a 
particular example of $f(\varphi)$.
In (\ref{action}) we have also introduced an arbitrary functional 
$P(\varphi , X)$, which can be expanded in power series in $X$ as 
\begin{eqnarray}
P(\varphi , X)=-V(\varphi)+
K(\varphi)X+L(\varphi )X^{2}+\cdots \cdots ~.
\label{eq:p}
\end{eqnarray}
The case of k-inflation \cite{Garriga:1999vw} corresponds 
to $V(\varphi)=0$. 
The non-canonical kinetic term of $\varphi $, which could be produced 
in underlying fundamental theories such as supergravity and superstring 
theories,  has many physical implications. We will, however, proceed 
without explicit use of the  expansion as in (\ref{eq:p})~.
In principle we could include  higher derivative terms 
of $\varphi$ as well. Effects of such higher derivative terms would be 
presumably suppressed by the ultraviolet cut-off scale. 
In so far as we consider the first derivative terms of $\varphi$, 
the action (\ref{action}) is the most general.

The equations of motion for the scalar field $\varphi $ and 
gravity turn out to be 
\begin{eqnarray}
\frac{1}{2}\frac{\partial f(\varphi)}{\partial \varphi}R+P,_{\varphi}+ \left (
 P,_{XX}\nabla ^{\mu}X+P,_{X\varphi }\nabla ^{\mu}\varphi 
 \right )\nabla _{\mu}\varphi +P,_{X}
\nabla _{\mu} \partial ^{\mu}
\varphi =0~, 
\label{eq:eomforvarphi}
\end{eqnarray}
\begin{eqnarray}
\nabla_{\lambda}\partial ^{\lambda}
f g_{\mu \nu}-\nabla _{\mu }\nabla _{\nu}f+fR_{\mu \nu}
-\frac{1}{2}fRg_{\mu \nu}
-
P,_{X}\nabla _{\mu}\varphi \nabla _{\nu}\varphi 
-
Pg_{\mu \nu}=0~.
\end{eqnarray}
The derivatives of $P$ with respect to $X$ and  $\varphi$ are denoted by 
$P_{,X}$ and $P_{,\varphi}$, respectively.
The classical solution $\varphi(t)$, which does not depend on 
spatial coordinates, is obtained by solving (\ref{eq:eomforvarphi}), and 
hereafter $\varphi $ shall always denote the classical solution after 
choosing an appropriate gauge (as will be discussed in 
 Eq. (\ref{eq:gaugecondition})).

\section{Conformal transformation to the Einstein frame}
\label{sec:conformal}

Given the general  action (\ref{action}), it is
useful to employ a conformal transformation to go to the Einstein frame,
which will be simpler and more familiar for us to develop 
cosmological perturbations. 
In the following we shall assume, in the Jordan frame, 
 the Friedmann-Lemaitre-Robertson-Walker (FLRW)   background
\begin{eqnarray}
{ds}^2=-dt^2+a^2(t)\delta_{ij}dx^i dx^j~.
\label{nmet0}
\end{eqnarray}

Now, let us consider the following conformal transformation:
\begin{eqnarray}
ds^{2} \to d\widehat{s}^{2}=
\Omega ^{2}ds^{2}
&=&
\Omega^{2} \left \{
g_{00}(dt)^{2}+2 \;g_{0i}\; dtdx^{i}
+g_{ij}dx^{i}dx^{j}\right \}
\nonumber \\
&=&
g_{00}(d\widehat{t})^{2}+2\;\Omega \;g_{0i}\; d\widehat{t}dx^{i}
+\Omega^{2}g_{ij}dx^{i}dx^{j}
\end{eqnarray}
where $i$ and $j$ are spatial indices and we have changed  
the time variable via the formula
\begin{eqnarray}
d\widehat{t}=\Omega dt =\Omega (\varphi(t))dt~.
\label{eq:dhattdtrelation}
\end{eqnarray} 
The functional form of $\Omega $ will be determined below, but 
note that $\Omega (\varphi(t))$ is a classical quantity, i.e., 
it does not contain a fluctuation part  (see Eq. (\ref{eq:gaugecondition})). 
We also assume $\Omega(\varphi (t))>0$ for the solution $\varphi(t)$. 
This assumption could be related with the stability of the ground state. 
Although the stability has been investigated in the literature 
\cite{Hosotani, futamasemaeda} for a particular 
choice of $f(\varphi)$ and is interesting in its own right, we  do not 
go into the details here.

We are thus led to redefine the metric in the following way:
\begin{eqnarray}
\widehat{g}_{00}=g_{00}, \hskip1cm
\widehat{g}_{0i}=\Omega g_{0i}, \hskip1cm
\widehat{g}_{ij}=\Omega^{2}g_{ij}~,
\hskip1cm
(i, j=1,2,3)~.
\label{eq:withandwithouthat}
\end{eqnarray}
The FLRW  background turns out in the Einstein frame to be 
\begin{eqnarray}
{d\widehat{s}}^2
=\Omega^{2} ds^{2}
=-d\hat{t}^2+\widehat{a}^2(\widehat{t})\delta_{ij}dx^i dx^j~.
\label{nmet}
\end{eqnarray}
where our notation is
\begin{eqnarray}
\widehat a(\widehat{t}) = \Omega (\varphi (t)) \, a(t)~,
\label{eq:ahatarelation}
\end{eqnarray}
and we should note again that $\widehat{a}(\widehat{t})$ does not 
contain a fluctuation part. 
Starting from the action   (\ref{action}) 
and using the conformal transformation
properties of the Ricci scalar \cite{Fujii:2003pa}, 
\cite{Wald:1984rg}
\begin{eqnarray}
R=\Omega^{2}\Bigg [
\widehat{R} + 6\widehat{\square} \ln \Omega -6\widehat{g}^{\mu\nu}
(\partial  _\mu\ln\Omega) (\partial _\nu\ln\Omega )
\Bigg ]~,
\label{eq:ricchi1}
\end{eqnarray}
we can show that the action (\ref{action}) is transformed as follows:
\begin{eqnarray}
S
&=& \frac 1 2\int d\widehat{t}\: d^{3}\vec{x}
\: \Omega^{-4}\sqrt{-\widehat{g}}
\nonumber \\
& & \times \Bigl[\Omega^2 f(\varphi)
 \Big\{ \widehat R + 6\widehat{\square} \ln \Omega 
 -6 \widehat{g}^{\mu\nu} (\partial _\mu\ln\Omega) (\partial _\nu\ln\Omega )
\Big \}+ 2P(\varphi,X) \Bigr].
\label{eq:ricchi2}
\end{eqnarray}
In order to arrive at the Einstein frame by making the coefficient 
of the scalar curvature unity,  we set
\begin{eqnarray}
\Omega (\varphi)=\sqrt{f(\varphi)}~.
\label{FW}
\end{eqnarray}
This puts the action into the following form: 
\begin{eqnarray}
S&=& \frac 1 2\int d\widehat{t} \: d^{3}\vec{x}\: \sqrt{-\widehat{g}}\Bigl[ 
\widehat{R}+2 \widehat{P}(\varphi,\widehat X) \Bigr]
\end{eqnarray}
where in the last step above the $\widehat \square$ term disappears
after integrating by parts 
and we have also defined 
\begin{eqnarray}
\widehat{X} &=& -\frac{1}{2}\widehat{g}^{\mu \nu}
\partial _{\mu} \varphi 
\partial _{\nu} \varphi 
=\frac{1}{\Omega^{2}} X~,
\\
\widehat{P}(\varphi,\widehat X)
&=&
\frac{1}{\Omega^{4}} P(\varphi, X) 
-3\widehat{g}^{\mu\nu} (\partial _\mu\ln\Omega) (\partial _\nu\ln\Omega )
\nonumber \\
&=&
\frac{1}{\Omega^{4}} P(\varphi, X) + \frac{6\widehat{X}}{\Omega ^{2}}
\left (\frac{d\Omega}{d\varphi}\right )^{2}~.
\label{eq:phatprelation}
\end{eqnarray}

The energy-momentum tensor of $\varphi $ is given in terms of 
$\widehat{P}(\varphi , \widehat{X})$ as 
\begin{eqnarray*}
{T^{0}}_{0}\equiv 
\widehat{E}
=2\widehat{X}\widehat{P}_{,\hat{X}}-\widehat{P}, 
\hskip1cm
{T^{i}}_{j}=-\widehat{P}{\delta ^{i}}_{j}, 
\hskip0.5cm (i, j=1,2,3).
\end{eqnarray*}
In our abbreviated notation, $\widehat{P}_{,\widehat{X}}$ is the derivative 
of $\widehat{P}$  with respect to $\widehat{X}$. 
The energy conservation law and the Friedmann equation 
are  given  as usual by
\begin{eqnarray}
\frac{d{\widehat{E}}}{d\widehat{t}}=-3\widehat{H}\left ( \widehat{E}+
\widehat{P} \right ), 
\hskip1cm
\widehat{H}^{2}\equiv 
\left ( \frac{1}{\widehat a}\frac{d\widehat a}{d\widehat{t}} 
\right )^2
=
\frac{\widehat{E}}{3}~.
\end{eqnarray}

\section{ADM formalism in the Einstein frame}
\label{sec: adm}


Cosmological perturbation theories have been developed for many years 
and the reader is referred to 
\cite{review} for  reviews.
In the case of the general quantum computation in gravity, we first
decompose the metric perturbations using the ADM formalism
\cite{Arnowitt:1962hi}, an application of which to cosmology 
has been done very early 
by Bardeen \cite{bardeen}. This method of decomposition has  
recently been applied to the study of non-Gaussianity 
by Maldecena \cite{maldacena} in standard
inflation and was then considered for more general models in
\cite{seerylidsey, chen1, Arroja:2008ga}.
Because we shall work in the Einstein frame we can simply use the
results already derived in \cite{seerylidsey, chen1, Arroja:2008ga} for
example, but with quantities now expressed with hats (\: $\widehat{}$\:\:). 
Thus, we shall
just briefly outline the method, which provides us with  a second order 
action  for the curvature perturbation (first derived by different
means in \cite{Garriga:1999vw}) and together with a third order 
action which leads
to the bispectrum  after applying the in-in formalism along 
lines similar to \cite{maldacena}.

We consider the metric which may be given as
\begin{eqnarray}
d\widehat{s}^{2}=-\widehat{N}^{2}(d\widehat{t}\:)^{2}
+\widehat{h}_{ij}\left ( dx^{i}+\widehat{N}^{i}d\widehat{t} \: \right )
\left ( dx^{j}+\widehat{N}^{j}d\widehat{t} \: \right ). 
\end{eqnarray}
The action in terms of the ADM variables is expressed as 
\begin{eqnarray}
& & S=\frac{1}{2}\int d\widehat{t}\:d^{3}\vec{x} \sqrt{\widehat{h}}
\widehat{N}\left \{
^{(3)}\widehat{R}+\frac{1}{\widehat{N}^{2}}\left ( 
\widehat{E}_{ij}\widehat{E}^{ij}-\widehat{E}^{2}\right )+ 
2\widehat{P}(\varphi , \widehat{X}) 
\right \}, 
\end{eqnarray}
where $\widehat{h}={\rm det}\widehat{h}_{ij}$, and the 
tensor $\widehat{E}_{ij}$ and $\widehat{E}$ are 
defined, respectively,  by
\begin{eqnarray}
\widehat{E}_{ij}=\frac{1}{2}\left (
\frac{\partial  \widehat{h}_{ij}}{\partial \widehat{t}}-
\widehat{\nabla }_{i}\widehat{N}_{j}-\widehat{\nabla }_{j}
\widehat{N}_{i}
\right ), 
\hskip0.5cm
\widehat{E}=\widehat{E}_{ij}\widehat{h}^{ij}~.
\end{eqnarray}
The covariant derivative $\widehat{\nabla}_{i}$ is with respect 
to $\widehat{h}_{ij}$. The three dimensional scalar curvature is 
denoted as $^{(3)}\widehat{R}$
%

We choose a gauge to fix time and spatial reparametrization 
by  setting  \cite{maldacena}
\begin{eqnarray}
\delta \varphi =0.
\label{eq:gaugecondition}
\end{eqnarray}
together with fluctuations around the FLRW metric  in the following form:
\begin{eqnarray}
d\widehat{s}^{2}=
-(d\widehat{t})^{2}+\widehat{a}(\widehat t)^{2}e^{2
\widehat{{\cal R}}} \left (
\delta _{ij}+\widehat{\gamma }_{ij}\right ), 
\hskip0.5cm 
(\partial ^{i}\widehat{\gamma} _{ij}=0, \:\: 
\delta^{ij}\widehat{\gamma }_{ij}=0)~.
\label{eq:fluctuation1}
\end{eqnarray}
Eqs. (\ref{eq:gaugecondition}) and (\ref{eq:fluctuation1}) fix the 
whole gauge degrees of freedom \cite{maldacena, nohhwang}.
Eq.  (\ref{eq:fluctuation1}) should be compared with the fluctuations in the Jordan frame
\begin{eqnarray}
ds^{2}
&=&
-(dt)^{2}+a(t)^{2} e^{2{\cal R} }\left ( 
\delta _{ij}+\gamma _{ij}\right )
\hskip1cm 
(\partial ^{i}\gamma _{ij}=0, \:\: \delta^{ij}\gamma _{ij}=0)~, 
\nonumber \\
&=&
\frac{1}{\Omega ^{2}}\left \{
-(d\widehat{t})^{2}+\widehat{a}(\widehat{t})^{2}
e^{2{\cal R}} \left ( \delta _{ij}+\gamma _{ij} \right )
\right \}~, 
\label{eq:fluctuation2}
\end{eqnarray}
where we have substituted  the relations (\ref{eq:dhattdtrelation}) and 
(\ref{eq:ahatarelation}).
Eqs. (\ref{eq:fluctuation1}) and (\ref{eq:fluctuation2}) indicate 
clearly that we need not distinguish ${\cal R}$ from 
$\widehat{{\cal R}}$  
or   $\gamma _{ij}$ from $\widehat{\gamma }_{ij}$
(as stressed in \cite{chiba, gong}), i.e., 
\begin{eqnarray}
\widehat{{\cal R}}={\cal R}, \hskip1cm \widehat{\gamma }_{ij}=\gamma _{ij}~.
\label{eq:basicformulae}
\end{eqnarray}
Hereafter we consider only the curvature perturbation 
$\widehat{{\cal R}}={\cal R}$,  and leave discussion of 
 the tensor perturbation $\widehat{\gamma }_{ij}=\gamma _{ij}$ 
 for future work.

The equations of motion for $\widehat{N}$ and $\widehat{N}^i$ 
are the Hamiltonian and
momentum constraints which are given by
\begin{eqnarray}
\widehat{\nabla }_{i}\left \{ \frac{1}{\widehat{N}}\left ( 
\widehat{E}^{i}_{j}-\delta ^{i}_{j}\widehat{E}
\right )\right \}=0, 
\label{eq:momentum}
\\
^{(3)}\widehat{R}+2\widehat{P}-4\widehat{X}\widehat{P}_{,\widehat{X}}
-\frac{1}{\widehat{N}^{2}}\left ( \widehat{E}_{ij}\widehat{E}^{ij}
-\widehat{E}^{2}\right )=0~.
\label{eq:hamiltonian}
\end{eqnarray}
We can solve the Hamiltonian and momentum constraints 
 order by order, by setting
\begin{eqnarray}
\widehat{N}=1+\widehat{N}^{(1)}+\cdots , 
\hskip1cm 
\widehat{N}_{i}=\partial _{i}\widehat{\psi }^{(1)}
+\widehat{\tilde N}^{(1)}_{i}
+\cdots ~,
\end{eqnarray}
where $\widehat{N}^{(1)}$,  $\widehat{\psi}^{(1)}$ and $\widehat{\tilde N}^{(1)}_i$ are 
the first order  and the ellipsis denotes higher order terms.
Here we assume $\partial ^{i}\widehat{\tilde N}^{(1)}_i=0$. The solutions to 
(\ref{eq:momentum})   and (\ref{eq:hamiltonian}) 
have been worked out \cite{seerylidsey, chen1} and are  known to be 
\begin{eqnarray}
\widehat{N}^{(1)}=\frac{1}{\widehat{H}}\frac{\partial 
\widehat{{\cal R}}}{\partial \widehat{t}},
\hskip0.5cm
\widehat{{\tilde N}}^{(1)}_{i}=0,
\hskip0.5cm
\widehat{\psi }^{(1)}=-\frac{\widehat{{\cal R}}}{\widehat{H}}
+\widehat{\chi }~,
\hskip0.5cm 
\left ( \widehat{H}=\frac{1}{\widehat{a}}
\frac{d\widehat{a}}{d\widehat{t}} \right )~.
\end{eqnarray}
Here $\widehat{\chi }$ is a solution to the  differential equation
\begin{eqnarray}
\partial ^{2}\widehat{\chi }=\widehat{a}^{2}\frac{\widehat{\varepsilon}}
{\widehat{c}_{s}^{2}}\frac{d\widehat{{\cal R}}}{d\widehat{t}}~,
\hskip2cm
\left ( \partial ^{2}=\delta ^{ij}\partial _{i}\partial _{j}\right ) ~, 
\end{eqnarray}
where the slow variation parameter and the sound speed are 
defined, respectively,  by
\begin{eqnarray}
\widehat{\varepsilon}=-\frac{1}{\widehat{H}^{2}}
\frac{d\widehat{H}}{d\widehat{t}}, 
\hskip1cm
\widehat{c}_{s}^{2}=\frac{\widehat{P}_{,\widehat{X}}}
{2\widehat{X}\widehat{P}_{,\widehat{X}\widehat{X}}+\widehat{P}_{,
\widehat{X}}}~.
\end{eqnarray}
As argued in \cite{maldacena,  chen1}, when calculating the effective 
action up to third order in $\widehat{\cal R}$, 
we need not calculate the $\widehat{\cal R}^{2}$ terms  
in $\widehat{N}$ or $\widehat{N}_{i}$.

Now let us turn to expansion of the action (\ref{action}) in power series 
of the curvature perturbation
\begin{eqnarray}
S=S^{(0)}+S^{(2)}+S^{(3)}+\cdots \cdots ~, 
\end{eqnarray}
where $S^{(n)}$ is the $n$-th order action.  Seery and Lidsey 
\cite{seerylidsey} and Chen et al. \cite{chen1} 
have studied $S^{(2)}$ and $S^{(3)}$  after performing integration by parts. 
After straightforward calculations they obtained
\begin{eqnarray}
S^{(2)}
&=&
\int d\widehat{t}\:d^{3}\vec{x}\Bigg \{
\widehat{a}^{3}\frac{\widehat{\varepsilon}}{\widehat{c}_{s}^{2}}
\left (
\frac{d\widehat{\cal{R}}}{d\widehat{t}}
\right )^{2}-\widehat{a}\widehat{\varepsilon}\left ( 
\partial \widehat{{\cal R}} \right )^{2}
\Bigg \}
\nonumber \\
&=&
\int d\widehat{t}\:d^{3}\vec{x}\Bigg \{
\widehat{a}^{3}\frac{\widehat{\Sigma}}{\widehat{H}^{2}}
\left (
\frac{d\widehat{\cal{R}}}{d\widehat{t}}
\right )^{2}-\widehat{a}\widehat{\varepsilon}\left ( 
\partial \widehat{{\cal R}} \right )^{2}
\Bigg \}~, 
\label{eq:quadraticactioninEinstein}
\end{eqnarray}
and 
\begin{eqnarray}
S^{(3)}&=&
\int d\widehat{t}\:d^{3}\vec{x}
\Bigg [
-\widehat{a}\widehat{\varepsilon}\widehat{{\cal R}}
(\partial \widehat{{\cal R}})^{2}
-\left ( \widehat{\Sigma}+2\widehat{\lambda} \right )
\frac{\widehat{a}^{3}}{\widehat{H}^{3}}\left ( \frac{d
\widehat{{\cal R}}}{d\widehat{t}}
 \right )^{3}+\frac{3\widehat{a}^{3}\widehat{\varepsilon} }
 {\widehat{c}_{s}^{2}}\widehat{{\cal R}} 
 \left ( \frac{d\widehat{{\cal R}}}{d\widehat{t}} 
 \right )^{2}
 \nonumber \\
 & &
 +\frac{1}{2\widehat{a}}\left ( 3\widehat{{\cal R}}
 -\frac{1}{\widehat{H}}\frac{
 d\widehat{{\cal R}}}{d\widehat{t}} \right )\left (
 \partial _{i}\partial _{j}\widehat{\psi}^{(1)} 
 \partial ^{i}\partial ^{j}\widehat{\psi}^{(1)} 
 - \partial ^{2}\widehat{\psi }^{(1)}
 \partial ^{2}\widehat{\psi }^{(1)}
 \right )
 \nonumber \\
 & &
 -\frac{2}{\widehat{a}}(\partial _{i}\widehat{\psi }^{(1)})
(\partial ^{i}\widehat{{\cal R}})
(\partial^{2}\widehat{\psi }^{(1)} ) \Bigg ]~,
\label{eq:cubicaction1}
\end{eqnarray}
which we have also confirmed.  Here we use notation analogous to that in 
\cite{seerylidsey} and \cite{chen1}, i.e.,
\begin{eqnarray}
\widehat{\Sigma}
&=&
\widehat{X}\widehat{P}_{,\widehat{X}}+
2\widehat{X}^{2}\widehat{P}_{,\widehat{X}\widehat{X}}
=
\frac{\widehat{X}\widehat{P}_{,\widehat{X}}}{\widehat{c}_{s}^{2}}
=
\frac{\widehat{\varepsilon} \widehat{H}^{2}}{\widehat{c}_{s}^{2}}~,
\label{eq:defofsigmahat}
\\
\widehat{\lambda} 
&=&
\widehat{X}^{2}\widehat{P}_{,\widehat{X}\widehat{X}}
+\frac{2}{3}\widehat{X}^{3}\widehat{P}_{,\widehat{X}\widehat{X}
\widehat{X}}~.
\label{eq:defoflambdahat}
\end{eqnarray}
Note that $\widehat{X}$, $\widehat{P}_{, \widehat{X}}$ and 
$\widehat{P}_{, \widehat{X}\widehat{X}}$ 
are all understood as classical values 
corresponding to $\varphi=\varphi (t)$. 
For possible operators that could appear in the third order action, 
see Ref. \cite{petel}.

\section{The Jordan frame}
\label{sec:jordan}


\subsection{The quadratic part of the action}

The definition (\ref{eq:defofsigmahat}) of $\widehat{\Sigma}$  
together with (\ref{eq:phatprelation}) enables us to express 
$\widehat{\Sigma}$ in terms of quantities without hat:
\begin{eqnarray}
\widehat{\Sigma}
&=&
\frac{1}{\Omega^{4}}
\left \{
XP_{,X}+2X^{2}P_{, XX}+6X\left ( \frac{d\Omega }{d\varphi } \right )^{2}
\right \}
\nonumber \\
&=&
\frac{1}{f(\varphi)^{2}}
\left \{
\Sigma +\frac{3}{4f(\varphi)}
\left ( \frac{df}{dt} \right )^{2}
\right \}
\nonumber \\
&=&
e^{-4\theta}a^{4}
\left \{ \Sigma + 3e^{2\theta}a^{-2}
\left ( \frac{d\theta}{dt} -H \right )^{2}
\right \}~.
\end{eqnarray}
Here we have introduced the notation 
\begin{eqnarray}
\Sigma =XP_{,X}+2X^{2}P_{, XX}~, 
\hskip0.5cm 
\theta =\frac{1}{2}{\rm log}(f(\varphi)a^{2}), 
\hskip0.5cm
H=\frac{1}{a}\frac{da}{dt}~. 
\end{eqnarray}
As we see $\Sigma $ is an analog in the Jordan frame,  corresponding to
$\widehat{\Sigma}$. 
The function $\theta$ has been employed by Qiu and Yang \cite{Qiu:2010dk}
 extensively. The Hubble constant defined in the Jordan frame is 
 denoted by $H$. 

The connection between  $\widehat{H}$  and the Hubble constant  $H$
in the Jordan frame  is easily seen from (\ref{eq:ahatarelation}) to be
\begin{eqnarray}
\widehat{H}
&=&
\frac{1}{\widehat{a}}\frac{d\widehat{a}}{d\widehat{t}}
=
\frac{H}{\Omega}+\frac{1}{\Omega ^{2}}\frac{d\Omega }{dt}
=
\frac{1}{\sqrt{f(\varphi)}}
\left \{
H+\frac{1}{2f(\varphi)}\frac{df(\varphi)}{dt}
\right \}
=
e^{-\theta}a\frac{d\theta}{dt}~.
\label{eq:hhatinjordan}
\end{eqnarray}
One of the standard parameters $\widehat{\varepsilon}$ in the Einstein frame 
can be expressed in terms of $\theta$ by using (\ref{eq:hhatinjordan}) as 
\begin{eqnarray}
\widehat{\varepsilon}
=
-\frac{1}{\widehat{H}^{2}}\frac{d\widehat{H}}{d\widehat{t}}
&=&
-\left ( \frac{d\theta}{dt} \right )^{-2}
\left \{
\frac{d^{2}\theta}{dt^{2}}+H\frac{d\theta}{dt}-\left ( 
\frac{d\theta}{dt} \right )^{2}
\right \}~.
\end{eqnarray}
The coefficients of each term in the quadratic action 
(\ref{eq:quadraticactioninEinstein})  can  all be 
expressed in terms of $\theta$, 
\begin{eqnarray}
\widehat{a}^{3}\frac{\widehat{\Sigma}}{\widehat{H}^{2}}
&=&e^{\theta}a^{2}
\left \{ \Sigma + 3e^{2\theta}a^{-2}
\left ( \frac{d\theta}{dt} -H \right )^{2}
\right \}\left ( \frac{d\theta}{dt} \right )^{-2}~, 
\label{eq:asigama/h}
\end{eqnarray}
\begin{eqnarray}
-\widehat{a}\widehat{\varepsilon}
=
e^{\theta}\left ( \frac{d\theta}{dt} \right )^{-2}
\left \{
\frac{d^{2}\theta}{dt^{2}}+H\frac{d\theta}{dt}-\left ( 
\frac{d\theta}{dt} \right )^{2}
\right \}~.
\label{eq:aepsilon}
\end{eqnarray}
We thus end up with the quardratic action 
\begin{eqnarray}
S^{(2)}
&=&
\int dt \: d^{3}\vec{x}\Bigg [
a^{3}
\left \{ \Sigma + 3e^{2\theta}a^{-2}
\left ( \frac{d\theta}{dt} -H \right )^{2}
\right \}\left ( \frac{d\theta}{dt} \right )^{-2}
\left ( \frac{d{\cal R}}{dt} \right )^{2}
\nonumber \\
& & 
\hskip1.5cm
+
e^{2\theta}a^{-1}
\left ( \frac{d\theta}{dt} \right )^{-2}
\left \{
\frac{d^{2}\theta}{dt^{2}}+H\frac{d\theta}{dt}-\left ( 
\frac{d\theta}{dt} \right )^{2}
\right \}
(\partial {\cal R})^{2}
\Bigg ]
\end{eqnarray}
in the Jordan frame. This agrees with the result obtained previously by
Qiu and Yang \cite{Qiu:2010dk} via a direct computation. 

\subsection{The cubic part of the action}

We would also like to express the cubic action (\ref{eq:cubicaction1}) 
by using quantities without hat. The definition (\ref{eq:defoflambdahat}) 
is shuffled into 
\begin{eqnarray}
\widehat{\lambda}
&=&\frac{1}{\Omega ^{4}}\left (
X^{2}P_{,XX}+\frac{2}{3}X^{3}P_{,XXX}
\right )
\equiv 
e^{-4\theta}a^{4}\lambda ~,
\end{eqnarray}
where $\lambda$, an analog in the Jordan frame, is defined by the 
 equation above.
The coefficient of the second  term in  the cubic action 
(\ref{eq:cubicaction1})  thus turns out to be
\begin{eqnarray}
\left (
\widehat{\Sigma}+2\widehat{\lambda }
\right )\frac{\widehat{a}^{3}}{\widehat{H}^{3}}
&=&
a\: e^{2\theta}\left \{
\Sigma + 2\lambda +3e^{2\theta}a^{-2}\left ( \frac{d\theta}{dt}-H \right )^{2}
\right \}
\left ( \frac{d\theta}{dt}\right )^{-3}~.
\end{eqnarray}
The quantities that appear in the third and fourth terms in 
(\ref{eq:cubicaction1}) are also worked out straightforwardly. 
In fact a combined use of (\ref{eq:defofsigmahat}) and (\ref{eq:asigama/h}) 
gives us 
\begin{eqnarray}
\frac{3\widehat{a}^{3}\varepsilon }{\widehat{c}_{s}^{2}}
&=&
\frac{3\widehat{a}^{3}\widehat{\Sigma}}{\widehat{H}^{2}}
\nonumber \\
&=&
3e^{\theta}a^{2}\left \{
\Sigma +3e^{2\theta}a^{-2}\left (
\frac{d\theta}{dt}-H
\right )^{2}
\right \}\left ( \frac{d\theta}{dt} \right )^{-2}
~.
\end{eqnarray}
From (\ref{eq:basicformulae}) and (\ref{eq:hhatinjordan}) , we also get 
\begin{eqnarray}
3\widehat{{\cal R}}-\frac{1}{\widehat{H}}\frac{d
\widehat{{\cal R}}}{d\widehat{t}}
=
3{\cal R}-\left (\frac{d\theta}{dt}\right )^{-1}
\frac{d{\cal R}}{dt}~.
\end{eqnarray}

We are thus led to the cubic action 
\begin{eqnarray}
S^{(3)}&=&\int dt\:d^{3}\vec{x}
\Bigg [
e^{2\theta}a^{-1}\left ( \frac{d\theta}{dt} \right )^{-2}\left \{ 
\frac{d^{2}\theta}{dt^{2}}+H\frac{d\theta}{dt}-\left ( 
\frac{d\theta}{dt} \right )^{2} \right \}
{\cal R}(\partial {\cal R})^{2}
\nonumber \\
& &-a^{3}
\left \{ \Sigma +2\lambda +3e^{2\theta}a^{-2}\left ( \frac{d\theta}{dt}
-H\right )^{2} \right \}\left ( \frac{d\theta}{dt} \right )^{-3}
\left ( \frac{d{\cal R}}{dt} \right )^{3}
\nonumber \\
& &
+3a^{3}\left \{
\Sigma +3e^{2\theta}a^{-2}\left (
\frac{d\theta}{dt}-H\right )^{2}
\right \}
\left ( \frac{d\theta}{dt} \right )^{-2}
 {\cal R}\left ( \frac{d{\cal R}}{dt} \right )^{2}
\nonumber \\
& &+
\frac{1}{2}e^{2\theta}a^{-3}\left \{ 3{\cal R}- \left ( \frac{d\theta}{dt} 
\right )^{-1}\frac{d{\cal R}}{dt} \right \}
\left (
\partial _{i}\partial _{j}\psi ^{(1)}\partial ^{i}\partial ^{j}\psi ^{(1)}
-\partial ^{2}\psi ^{(1)}\partial ^{2}\psi ^{(1)}
\right )
\nonumber \\
& &-2e^{2\theta}a^{-3}(\partial _{i}\psi ^{(1)})(\partial ^{i}{\cal R})
(\partial ^{2}\psi ^{(1)})
\Bigg ]~, 
\label{eq:cubic-in-Jordan}
\end{eqnarray}
which is suitable for computation of cosmological correlations in the 
Jordan frame. In the above we have set 
\begin{eqnarray}
\widehat{\psi}^{(1)}=\Omega \psi ^{(1)}=e^{\theta}a^{-1}\psi ^{(1)}~.
\label{eq:psi}
\end{eqnarray}
Note that Eq. (\ref{eq:psi}) is due to the fact that the lapse 
function, shift function and the three-dimensional metric  in 
Einstein and Jordan frames are connected by the formulae
\begin{eqnarray}
\widehat{N}=N, 
\hskip0.5cm 
\widehat{N}_{i}=\Omega N_{i}, 
\hskip0.5cm 
\widehat{N}^{i}=\frac{1}{\Omega }N^{i}, 
\hskip0.5cm
\widehat{h}_{ij}=\Omega^{2}h_{ij}~, 
\end{eqnarray}
in accordance with (\ref{eq:withandwithouthat}).{
The cubic action (\ref{eq:cubic-in-Jordan}) agrees with   
the result  in \cite{Qiu:2010dk} obtained by a direct method.

\section{Frame independence of quantum calculations}
\label{sec:frame}

As a final  step of our analyses let us scrutinize the properties of 
quantization in the Einstein and Jordan frames. 
The Lagrangian densities in Einstein and Jordan frames are 
connected by the relation $\Omega \widehat{{\cal L}}={\cal L}$. This 
can be seen by rewriting the action in the following way:
\begin{eqnarray}
S=\int d\widehat{t}\:d^{3}\vec{x}\:\widehat{{\cal L}}
=
\int dt\:d^{3}\vec{x}\:\Omega \widehat{{\cal L}}
=
\int dt\:d^{3}\vec{x}\:{\cal L}~.
\end{eqnarray}
The canonical momentum variables conjugate to $\widehat{{\cal R}}$ and 
${\cal R}$, i.e., 
\begin{eqnarray}
\widehat{\Pi} =\frac{\delta \widehat{{\cal L}}}{\delta 
(\partial \widehat{{\cal R}}/\partial \widehat{t}\:)}, 
\hskip1cm
\Pi =\frac{\delta {\cal L}}{\delta 
(\partial {\cal R}/\partial t)}
\end{eqnarray}
are also apparently indentical with each other, namely,
\begin{eqnarray}
\widehat{\Pi}(\widehat{t}, \vec{x})=\Pi (t, \vec{x})~.
\end{eqnarray}
The canonical commutation relations 
\begin{eqnarray}
\left [ \widehat{{\cal R}}(\widehat{t}, \vec{x}), 
\widehat{\Pi }(\widehat{t}, \vec{x}') \right ]=i\:
\delta^{3}(\vec{x}-\vec{x}')~, 
\hskip1cm
\left [ {\cal R}(t, \vec{x}), \Pi (t, \vec{x}') \right ]=
i\:\delta^{3}(\vec{x}-\vec{x}')
\end{eqnarray}
lead to  the same  quantization procedure common 
in both Einstein and Jordan frames.

Let us turn to the mode expansion in Einstein and Jordan frames, 
\begin{eqnarray}
\widehat{{\cal R}}(\widehat{t}, \vec{x})
&=&
\int \frac{d^{3}\vec{k}}{(2\pi )^{3}}
\left \{
\widehat{a}(\vec{k})\widehat{u}(\widehat{t}, \vec{k})e^{i\vec{k}\cdot \vec{x}}
+
\widehat{a}^{\dag}(\vec{k})\widehat{u}^{*}(\widehat{t}, \vec{k})
e^{-i\vec{k}\cdot \vec{x}}\right \}~,
\\
{\cal R}(t, \vec{x})
&=&
\int \frac{d^{3}\vec{k}}{(2\pi )^{3}}
\left \{
a(\vec{k})u(t, \vec{k})e^{i\vec{k}\cdot \vec{x}}
+
a^{\dag}(\vec{k})u^{*}(t, \vec{k})
e^{-i\vec{k}\cdot \vec{x}}\right \}~,
\end{eqnarray}
and look at the connection between creation and annihilation operators 
in the two frames. Hereafter all quantities are supposed to be 
in the interaction picture. The differential 
equations to be satisfied by the mode functions are given 
by the quadratic part of the action and are 
\begin{eqnarray}
\left \{
\frac{d}{d\widehat{t}}\left ( 
\widehat{a}^{3}\frac{\widehat{\Sigma}}{\widehat{H}^{2}} 
\frac{d}{d\widehat{t}}\right ) 
+\widehat{a}\widehat{\varepsilon}\: \vec{k}^{2} 
\right \}\widehat{u}(\widehat{t}, \vec{k})=0
\label{eq:modefunction1}
\end{eqnarray}
and 
\begin{eqnarray}
& &
\frac{d}{dt}\left [
a^{3}\left \{ \Sigma + 3e^{2\theta}a^{-2}
\left ( \frac{d\theta}{dt} -H \right )^{2}
\right \}\left ( \frac{d\theta}{dt} \right )^{-2}
 \frac{d}{dt}
 \right ]u(t, \vec{k})
 \nonumber \\
& &
\hskip1cm -
e^{2\theta}a^{-1}
\left ( \frac{d\theta}{dt} \right )^{-2}
\left \{
\frac{d^{2}\theta}{dt^{2}}+H\frac{d\theta}{dt}-\left ( 
\frac{d\theta}{dt} \right )^{2}
\right \}\vec{k}^{2}u(t, \vec{k})
=0
\label{eq:modefunction2}
\end{eqnarray}
in the two frames, respectively. 
Qiu and Yang \cite{Qiu:2010dk} elaborated the solutions to 
(\ref{eq:modefunction2}) in the Jordan frame. 
It should be,  however,  remarked  that Eq. (\ref{eq:modefunction2}) 
can be  equivalently rewritten as
\begin{eqnarray}
\left \{
\frac{1}{\Omega}\frac{d}{dt}\left (
\widehat{a}^{3}\frac{\widehat{\Sigma}}{\widehat{H}^{3}}\frac{1}{\Omega}
\frac{d}{dt}\right )
+\widehat{a}\widehat{\varepsilon}\:\vec{k}^{2}
\right \} u(t, \vec{k})=0~, 
\label{eq:modefunction3}
\end{eqnarray}
thanks to the relation (\ref{eq:asigama/h}) and (\ref{eq:aepsilon}). 
Furthermore, it is also obvious that 
(\ref{eq:modefunction3}) is the same differential equation 
as (\ref{eq:modefunction1}) because of (\ref{eq:dhattdtrelation}). 

The solutions to (\ref{eq:modefunction2}) can be worked out by 
going over to the Einstein frame, thereby making full use of 
 results obtained previously. In fact,  
by adjusting the normalization of $u(t, \vec{k})$ and 
$\widehat{u}(\widehat{t}, \vec{k})$ in the same manner, we can always set 
\begin{eqnarray}
& & \widehat{u}(\widehat{t}, \vec{x})=u(t, \vec{x})~, 
\\
& &
\widehat{a}(\vec{k})=a(\vec{k}), 
\hskip2cm
\widehat{a}^{\dag}(\vec{k})=a^{\dag}(\vec{k})~.
\label{eq:creationandannihilation}
\end{eqnarray}
The ``free vacuum" $\vert 0 \rangle $ is defined as  the state that 
is annihilated by all of 
the annihilation operators, i.e., 
$
a(\vec{k})\vert 0 \rangle =0, 
$
and 
$
\widehat{a}(\vec{k})\vert 0 \rangle =0~.
$
Because of (\ref{eq:creationandannihilation}), 
the free vacuum $\vert 0 \rangle $ is common in both Einstein 
and Jordan frames.

In the interaction picture, the time-development of the 
``interaction  vacuum" 
starting from $\vert 0 \rangle $ at some early time ($\widehat{t}_{0}$  or 
$t_{0}$) 
is described by the unitary transformation composed of the 
interaction Hamiltonian, that is, 
\begin{eqnarray}
\vert \widehat{{\rm vac}} \rangle =
T {\rm exp}\left (-i\:\int _{\widehat{t}_{0}}^{\widehat{t}}
\:d\widehat{t'} \:\widehat{H}_{\rm int}(\widehat{t'})\right )
\vert 0\rangle
\label{eq:vacuum-einstein}
\end{eqnarray}
in the Einstein frame and 
\begin{eqnarray}
\vert {\rm vac} \rangle = 
T {\rm exp}\left (-i\:\int _{t_{0}}^{t}\:dt' \:H_{\rm int}(t')\right )
\vert 0\rangle  
\label{eq:vacuum-jordan}
\end{eqnarray}
in the Jordan frame. 
Here $T$ indicates the time-ordering and the interaction Hamiltonian is 
denoted by  $\widehat{H}_{\rm int}(\widehat{t})$ 
($H_{\rm int}(t)$) in the Einstein (Jordan) frame. 
Now we show that the two states, (\ref{eq:vacuum-einstein}) and 
(\ref{eq:vacuum-jordan}),  are in fact identical.
From the preceding analyses, we can easily 
see that $\widehat{H}_{\rm int}(\widehat{t})$ is connected 
with  $H_{\rm int}(t)$  via
\begin{eqnarray}
\widehat{H}_{\rm int}(\widehat{t})=\frac{1}{\Omega(\varphi(t))}
H_{\rm int}(t)~.
\label{eq:interactionhamiltonian}
\end{eqnarray}
Combining (\ref{eq:interactionhamiltonian}) with 
(\ref{eq:dhattdtrelation}), we obtain a simple relation: 
\begin{eqnarray}
\int _{\widehat{t}_{0}}^{\widehat{t}}\: d\widehat{t'} \: 
\widehat{H}_{\rm int}(\widehat{t'})
=
\int _{t_{0}}^{t}\: dt' \: H_{\rm int}(t')~,
\label{eq:timedevelopment}
\end{eqnarray}
and therefore the two vacua are identical, i.e., 
$\vert \widehat{{\rm vac}} \rangle =\vert {\rm vac} \rangle $.

In Sec. \ref{sec:jordan}  we  discussed only the quadratic and cubic terms,  
$S^{(2)}$ and $S^{(3)}$ in the action, but our argument 
can  obviously be extended to higher order terms 
in $\widehat{{\cal R}}={\cal R}$. 
Since the interaction vacuum is one and the same in  both frames, the 
cosmological correlation functions computed in each frame can be shown to 
 agree with each other,  that is, 
\begin{eqnarray}
\langle \widehat{{\rm vac}} \vert \widehat{{\cal R}}(\widehat{t}, \vec{k}_{1})
\cdots 
\widehat{{\cal R}}(\widehat{t}, \vec{k}_{N}) \vert \widehat{{\rm vac}} \rangle
&=&
\langle {\rm vac} \vert {\cal R}(t, \vec{k}_{1})
\cdots 
{\cal R}(t, \vec{k}_{N}) \vert {\rm vac} \rangle
\label{eq:generalNrelation}
\end{eqnarray}
for  an arbitrary integer $N$, not to mention the $N=2$ 
(power spectrum) and $N=3$ (bispectrum) cases.
Here  ${\cal R}(t, \vec{k})$ 
($\widehat{{\cal R}}(\widehat{t}, \vec{k})$) is the 
three-dimensional Fourier transform 
of ${\cal R}$ ($\widehat{\cal R}$) in the Jordan (Einstein) frame.

\section{Summary}
\label{sec:summary}

In the present paper we have made an overall comparison between two types 
of cosmological perturbation theories, one in the Einstein and the other 
in the Jordan frame. By virtue of the basic relation 
(\ref{eq:basicformulae}), the connection between the two frames 
is  extremely simple 
and we can easily go back and forth between the two frames. 
In this way we are able to find a way for alleviating the need 
for separate studies in the two frames.

In the actual calculation of the correlation (\ref{eq:generalNrelation}), 
we have to evaluate multiple-commutators  \cite{weinberg}
\begin{eqnarray}
& &
\hskip-1cm
\langle \widehat{{\rm vac}} \vert 
\widehat{{\cal R}}(\widehat{t}, \vec{k}_{1})
\cdots 
\widehat{{\cal R}}(\widehat{t}, \vec{k}_{N}) 
\vert \widehat{{\rm vac}} \rangle
\nonumber \\
&=&
\sum _{n=0}^{\infty}
i^{n}\int _{\widehat{t}_{0}}^{\widehat{t}}\:d\widehat{t}_{n}
\int _{\widehat{t}_{0}}^{\widehat{t}_{n}}\:d\widehat{t}_{n-1}
\cdots \cdots
\int _{\widehat{t}_{0}}^{\widehat{t}_{2}}\:d\widehat{t}_{1}
\nonumber \\
& & \times \langle 0 \vert [\widehat{H}_{\rm int}(\widehat{t}_{n}), [
\widehat{H}_{\rm int}(\widehat{t}_{n-1}), 
\cdots \cdots [
\widehat{H}_{\rm int}(\widehat{t}_{1}), 
\widehat{{\cal R}}(\widehat{t}, \vec{k}_{1})
\cdots 
\widehat{{\cal R}}(\widehat{t}, \vec{k}_{N}) 
 ]\cdots \cdots ]] 
\vert 0 \rangle ~
\end{eqnarray}
in the Einstein frame or similarly in the Jordan frame.
Thanks to  (\ref{eq:timedevelopment}) 
every step of calculating the commutators in the Einstein and Jordan 
frames may be  compared  with each other 
and can be shown  to be identical.

\vskip1cm
\noindent
{\bf Acknowledgements}

The authors would like to express their sincere thanks 
to Professor Tetsuya Onogi for collaboration in the early stages 
of the present work. 
They are also grateful to Professor Misao Sasaki and Dr. 
Masato Minamitsuji for useful discussions. The authors' cordial 
thanks should also go to Professor Takeshi Chiba and Professor 
Masahide Yamaguchi for calling their attention to Ref. \cite{chiba}.
\vskip1cm
\noindent
Note added : After sending the present paper to the arXiv, 
we were informed by Dr. Yun-Song Piao of his recent work \cite
{piao}. He has shown that the primordial curvature perturbation and 
tensor perturbation generated during two conformally related 
backgrounds are equal. We were also informed by Dr. Sandipan 
Kundu of his recent work \cite{kundu}, in which the behavior of the 
curvature perturbation under the three-dimensional rescaling and 
its implications to non-Gaussianity are discussed. We would like to 
thank Dr. Yun-Song Piao and Dr. Sandipan Kundu for their kind 
communication.

\end{document}